\documentclass[aps,pre,twocolumn,float]{revtex4}
\usepackage{amsmath,bm,epsfig}

\def\Fbox#1{\vskip1ex\hbox to 8.5cm{\hfil\fboxsep0.3cm\fbox{%
  \parbox{8.0cm}{#1}}\hfil}\vskip1ex\noindent}  

\let \= \equiv \let\*\cdot \let\~\widetilde \let\^\widehat \let\-\overline

\newcommand {\ms}{\langle \sigma\rangle}

\begin{document}
\title{Locality and Non-locality in Elasto-plastic Responses of Amorphous Solids}
\author{Edan Lerner and Itamar Procaccia}
\affiliation{Department of Chemical Physics, The Weizmann
Institute of Science, Rehovot 76100, Israel  }
\date{\today}
\begin{abstract}
A number of current theories of plasticity in amorphous solids assume at their basis that plastic deformations are
spatially localized. We present in this paper a series of numerical
experiments to test the degree of locality of plastic deformation. These experiments increase in terms of the stringency
of the removal of elastic contributions to the observed elasto-plastic deformations. It is concluded that for all our simulational protocols the plastic deformations are {\bf not} localized, and their scaling is sub-extensive. We offer a number of measures of the magnitude of the plastic deformation, all of which display sub-extensive scaling characterized by non-trivial exponents.
We provide some evidence that the scaling exponents governing the sub-extensive scaling laws are non-universal, depending
on the degree of disorder and on the parameters of the systems. Nevertheless understanding what determines these exponents should shed considerable light on the physics of amorphous solids.
\end{abstract}
\maketitle

\section{Introduction}

\subsection{Motivation}

Much of the theoretical analysis of deformation and plastic flow in amorphous solids and other non-crystalline materials (structural glasses, metallic glasses, pastes, foams, gels etc.) is still influenced to a large degree by the understanding of crystalline materials. In the latter the deformation and flow are governed by topological defects known as dislocations, whose dynamics are the basis of the theory of crystalline plasticity \cite{66ST}. Indeed, following the pioneering work of Maeda and Takeuchi \cite{80KMT,81MK}
on metallic glasses and those of Argon and coworkers \cite{79AK,82AS} on bubble rafts, workers in the field of amorphous elasto-plasticity \cite{79Arg,98FL} proposed theoretical schemes based on the notion that also in amorphous solids plastic events are carried by some sort of micro-structural defects, referred to as ``Shear Transformation Zones" (STZ). While the precise nature of these STZ or how to measure them experimentally or even simulationally has never been fully clarified, their existence as the source of `quanta' of plastic relaxation carried by a small number of atoms was taken as a basis for developing mean-field models of elasto-plasticity. Although these models are not unique and are sometimes even in disagreement with each other, careful attempts to apply them to a variety of phenomena, from shear banding \cite{07MLC} to fracture \cite{06BPP} and from necking instabilities \cite{03ELP} to grooving via Grinfeld instabilities \cite{08LPPZ} all showed considerable promise and a fair agreement with experiments or simulations. The fundamental question of whether plasticity in amorphous solids is indeed due to local events in which a microscopic number of atoms (independent of the system size) are involved remained unanswered.

A serious doubt on this fundamental assertion was recently cast by Maloney and Lema$\^i$tre \cite{06ML} with their presentation of a series of two-dimensional atomistic computer simulations of amorphous solids subject to simple shear in the athermal, quasistatic limit. These authors argued that the plastic events themselves were lines of slip which span the length of the simulations cell. If it were shown that these findings extend to physical experiments, this would put in question the very fundamental assumptions underlying mean-field theories that were put forward, irrespective of their relative success to parameterize a number of interesting elasto-plastic phenomena. In a recent meeting at the Lorentz center in Leiden the conclusions of Maloney and Lema$\^i$tre were criticized \cite{Lan} on the basis of the algorithm used, in which after each step of strain the energy was minimized, irrespective of the computer time needed for this minimization or whether the trajectory followed is physical. This opens up the possibility that the spanning events seen in \cite{06ML} would not be seen in a system with `natural' dynamics in which the limit of zero strain rate is not to be confused with arbitrary waiting times between strain steps.

The aim of this paper is not to form judgment about the assumed locality of plastic deformations in laboratory experiments done at finite strain rates and finite temperatures. Rather, we focus on the fundamental issue of the spatial extent of plastic deformations using a variety of algorithms and a number of simulational tests for the locality (or rather non-locality) of these deformations in two-dimensional amorphous solids. Among other things we provide evidence that the results found in \cite{06ML} are generic. But we go further in analyzing the system-size dependence of various measures of elasto-plastic deformation. Our conclusions are even more pessimistic than those given by \cite{06ML}; we conclude that it is quite difficult to try to isolate the plastic contribution to an elasto-plastic event. The `plastic energy' part of such an event is ill-defined; every plastic deformation is accompanied by an elastic deformation which is trivially long ranged because elasticity is long ranged. The elastic contribution to every energy change in an elasto-plastic event can be much larger than the purely plastic energy change even if the latter were well defined. Measuring the latter is almost like weighing the captain by weighing the captain with the ship and and ship without the captain and taking the difference. We therefore propose below a number of new measures that are able to distinguish the irreversible plastic contribution from the elastic affine and non-affine contributions. Our conclusion is that the generic plastic deformations are not localized, and this is {\bf not} because of the elastic response that accompanies it. We will show that direct measures of the `size' of the purely plastic events scale with the size of the system similarly to the scaling of the elasto-plastic events.

The structure of this paper is as follows. In the second part of the introduction we describe the model used in the majority
of the paper below. In Sect. \ref{pullin} we describe an experiment of pulling one particle in a glassy system of varying sizes and measuring the maximal force on this particle before the plastic deformation. We show that the average maximal force depends on the size of the system as a power law, indicating that the system is never too large such that the walls do not matter. The power law
has a non-trivial exponent that we discuss below. In Sect. \ref{drops} we consider systems of various sizes under strain,
and study the scaling properties of the distributions of stress and energy drops when the system reaches a steady-state plastic flow. We demonstrate that these distributions are characterized by sub-extensive scaling with non-trivial exponents, again
indicating that the plastic events are not localized. We introduce a measure of the size of plastic events that filters out
the effect of non-affine elasticity, and show that this measure scales with essentially the same exponent as the totality
of the elasto-plastic energy drop. Finally, to remove the last doubts, in Sect. \ref{another} we introduce a new model glass
for which we can precisely measure the size and extent of a purely plastic event, filtering out completely any possible
elastic contribution. The conclusion is as above, that the size of plastic events scales in a sub-extensive fashion with a non-trivial exponent. In the discussion section \ref{summary} we provide a summary of the result and discuss the apparent
non-universality of the scaling exponents. It is stated that understanding the numerical values of the found exponents should be
an important step in improving our understanding of the physics of amorphous solids.

\subsection{Model Description}
\label{model}
\begin{figure}
\hskip -1.2cm
\centering
\includegraphics[scale = 0.55]{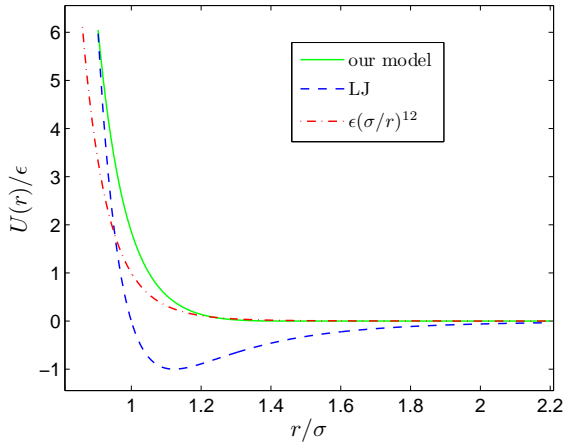}
\caption{The potential Eq. (\ref{potential}) used in the present simulations in comparison to the more standard $\sigma/r^{12}$ potential and to the Lennard-Jones potential. }
\label{potent}
\end{figure}

Almost all the simulations below are performed in a glassy system consisting of poly-dispersed soft disks.
We work with point particles of equal mass $m$ in two dimensions with pair-wise interaction potentials. Each particle $i$ is assigned a interaction parameter $\sigma_i$ from a normal distribution with mean 1. The variance is governed by the poly-dispersity parameter $\Delta = 15\%$ where $\Delta^2 = \frac{\langle \left(\sigma_i -
\langle \sigma \rangle \right)^2\rangle}{\langle \sigma \rangle^2} $. With the definition $\sigma_{ij} = \case{1}{2}(\sigma_i + \sigma_j)$, the potential assumes the form (cf. Fig. \ref{potent})
\begin{widetext}
\begin{equation}
U(r_{ij}) =
\left\{
\begin{array}{ccl}
\!\!\! \epsilon\left[\left(\frac{\sigma_{ij}}{r_{ij}}\right)^{k}\!\! -\!\!\frac{k(k+2)}{8}
\left( \frac{B_0}{k} \right)^{\frac{k+4}{k+2}}\left(\frac{r_{ij}}{\sigma_{ij}}\right)^4
+ \frac{B_0(k+4)}{4}\left(\frac{r_{ij}}{\sigma_{ij}}\right)^2
-\frac{(k+2)(k+4)}{8}\left( \frac{B_0}{n} \right)^{\frac{k}{k+2}}\right] & , & r_{ij} \le
\sigma_{ij}\left( \frac{k}{B_0} \right)^{\frac{1}{k+2}} \\
0 & , & r_{ij} >
\sigma_{ij}\left( \frac{k}{B_0} \right)^{\frac{1}{k+2}}
\label{potential}
\end{array}
\right\}\ ,
\end{equation}
\end{widetext}
Below the units of length, energy, mass and temperature are $\langle \sigma\rangle$, $\epsilon$, $m$ and $\epsilon/k_B$ where
$k_B$ is Boltzmann's constant. The time units $\tau_0$ are accordingly $\tau_0=\sqrt{(m\langle \sigma\rangle^2/\epsilon})$.
The motivation of this somewhat lengthy form of the potential is to have continuous first and second derivatives at the built-in cutoff of $ r_{ij} = \sigma_{ij}\left(k/B_0 \right)^{\frac{1}{k+2}}$. In the present simulations we chose $k=10$, $B_0=0.2$. The choice of a quartic rather than a quadratic correction term is motivated by numerical speed considerations, avoiding
the calculation of square roots. In all the simulations discussed below the number density $N/V=1.176$ and the
boundary conditions are periodic.

\section{System-size dependence of the force on a single particle}
\label{pullin}

\begin{figure}
\centering
\includegraphics[scale = 0.4]{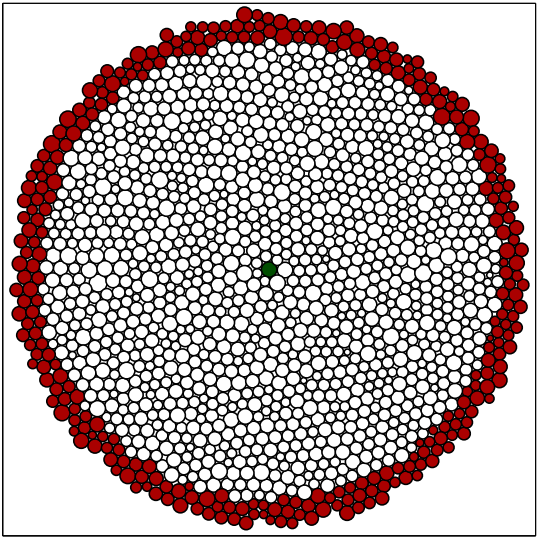}
\caption{Color online. The configuration of the pulling experiment. The center particle (in dark green) is pulled in a random direction and the force exerted on it is measured, in addition to the total energy of the system. The outer particles (in red) are stationary and are not allowed to move. We are interested in influence of the radius $R$ on the measured quantities.}
\label{showcircle}
\end{figure}

To initiate the discussion of locality vs. non-locality issues we begin with a simulation of the lovely experiment
presented in \cite{09Ban}. In this experiment one pulls a {\bf single} disk at a constant (low) velocity through
a disordered array of disks whose diameters have two possible values. In our case we use the system described in Subsect. \ref{model}. We start by quenching a system of $N=16384$ particles from $T=1.0$ to $T=0.05$ at a cooling rate of $10^{-6}$ $\epsilon/(k_B\tau_0)$. At this point we remove any residual heat by conjugate gradient minimization \cite{website}.
After this preparation we choose one particle out of the $N$ available ones (referred to below as the `center' particle), and draw a circle of radius $R$ around it. Next we freeze all the particles outside this circle, leaving the particles inside the circle to interact normally according to Newton's equations. An example of the results of such a procedure is shown in Fig. \ref{showcircle}. The procedure is then repeated for each and every one of the $N$ particles for the sake of statistics.

\begin{figure}
\hskip -0.5 cm
\centering
\includegraphics[scale = 0.53]{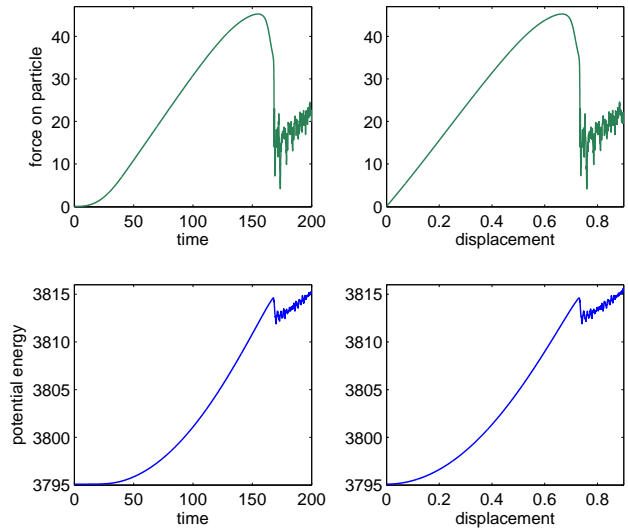}
\caption{Color online. The force exerted on the center particle and the potential energy of the system as a function of time (left panels) and as a function of displacement (right panels) for a typical run with $R=18.0$. Note the linearity of the force vs. displacement before the plastic failure, this is the elastic branch which is linear up to the maximum. Sometime non-affine elasticity destroys the linearity seen here.  At zero time the velocity increases smoothly from zero (up to second derivative)
in order to minimize shock waves.}
\label{trajpul}
\end{figure}

The experiment performed consists of pulling the center particle at a velocity $v_0=0.005~ \ms/ \tau_0$. We reach this velocity with a smooth initiation as seen in Fig. \ref{trajpul} in order to minimize elastic shock waves. Obviously, when the particle
begins to move it increases the elastic energy in the whole system. This increase continues until the first plastic event in which
an irreversible re-organization of the particle positions takes place. This event is irreversible in the sense that until it happens one can reverse the motion and return to the initial condition, but after the event heat is released in the form of kinetic energy and reversibility is lost. An example of a typical run is available in a movie that can be found in \cite{web}, and the
results of this run are displayed in Fig. \ref{trajpul}. One can see in the movie that the plastic event appears local to the eye.
The main question of this paper is whether this is just an eyeball impression or is it quite true. We study this issue by changing the radius $R$ and examining the distribution of the maximal force exerted on the particle before the plastic failure. At each value of $R$ we repeat the experiment $N$ times, once for each particle in the system being the center particle.  In each trial the center particle is pulled in a random direction and we measure the force exerted on it in the direction of the motion.

\begin{figure}
\centering
\hskip -.5 cm
\includegraphics[scale = 0.40]{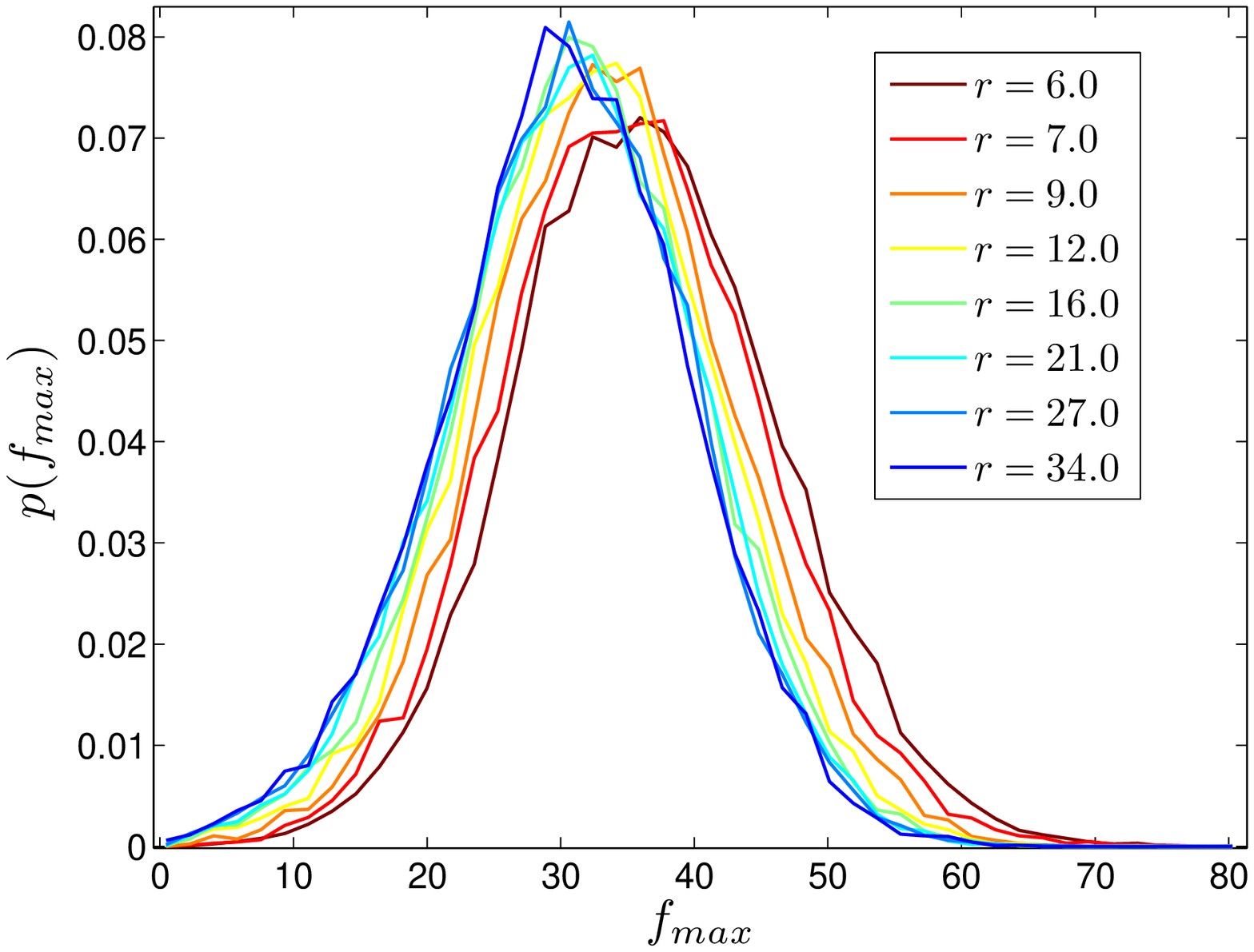}
\includegraphics[scale = 0.45]{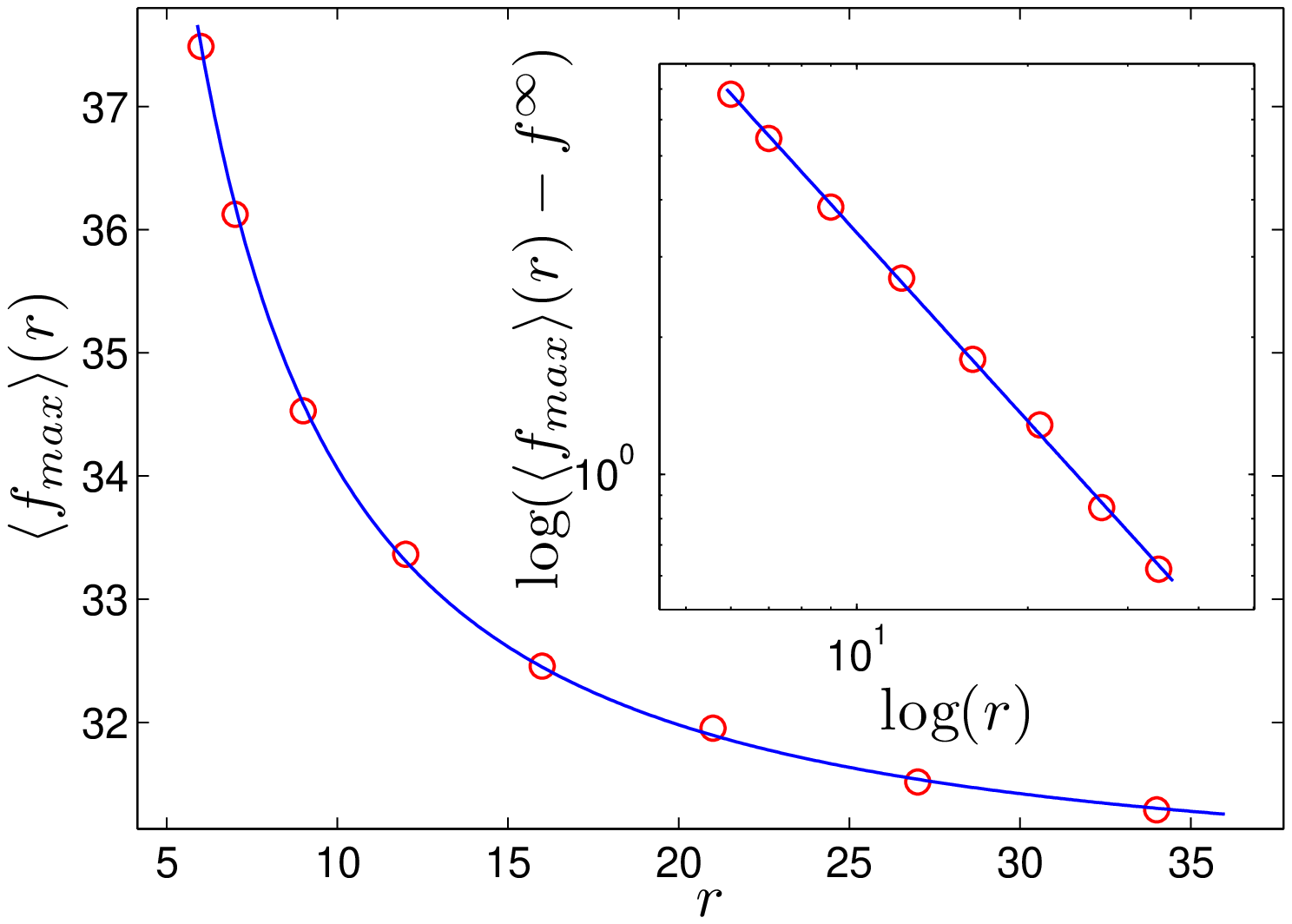}
\caption{Upper panel: the distributions of maximal force before plastic event in the pulling experiment, for different radii
of system sizes. The radius increases from right to left. Lower panel: the scaling of the mean maximal force as a function of the
system size, demonstrating the scaling law Eq. (\ref{slaw}) in linear coordinates and in the inset as a log-log plot.}
\label{maxforce}
\end{figure}

In Fig. \ref{trajpul} one can see the force as a function of time and as a function of displacement for a typical run. Naively one could expect that if the plastic failure were a localized phenomenon, then the average (over $N$) maximal force in the direction of pulling before that failure should not depend crucially on the circle radius $R$, or if it does depend on $R$ that dependence should fall off exponentially to an $R$-independent value as $R$ increases. To test this expectation we compute the average maximal force for different values of $R$. In fact, when we increase $R$ we find that the maximal force falls off very slowly as a function of $R$, as a power law, cf. Fig. \ref{maxforce}.
The distributions of maximal force, $p(f_{\rm max})$ moves systematically to smaller values of $f_{\rm max}$ when $R$ increases. For the present system we find a power law decay in the average value of the maximal force, with the excellent fit exhibited in
the lower panel of Fig. \ref{maxforce}. The power law reads
\begin{equation}
\langle f_{\rm max} \rangle = f^\infty + B R^{-\gamma} \ ,
\label{slaw}
\end{equation}
with $B=79.84$, $\gamma=1.37$ and $f^\infty=30.67$ being the asymptotic average maximal force for $R\to \infty$. This power law
is the first one found in this paper with a non-trivial exponent, others will follow below. We have checked that the exponent in the power law is {\em not} universal, being dependent on the potential and other characteristics of the system. At this point we do not have a solid theory to predict the value of either this or the other exponents discussed below, but we conjecture that the present exponent is determined by the fractal dimension $D$ of the force-chains created by the loading, and may even be precisely that number, i.e $\gamma=D$. Obviously it is highly desirable to develop such scaling relations in the near future.

The reason of the slow decay in the average maximal force is that in disordered systems one can always find regions that
either respond via non-affine elasticity or yield plastically for smaller forces when the system increases in size and the availability of softer regions becomes apparent. Nevertheless the convergence to a finite average maximal force for $R\to \infty$ shows that there exists a material parameter, analogous to the yield stress, which determines the density of weak pathways per unit volume when the system is sufficiently large. We can therefore conclude that the plastic events incurred during the pulling of the center particle `know' about the size of the system, and the disorder does not screen the boundary from the local loading. One could argue that this is expected since the loading creates force chains that must end at the boundaries, and that the long range effects found here are nothing but a consequence of the fact that elasticity is long ranged. Indeed, it is very difficult
to disentangle purely plastic from elastic contributions in any elasto-plastic material. Thus to drive our point further
we turn next to the stress and energy drops during plastic irreversible deformations under a different kind of loading.

\section{System-size dependence of elasto-plastic events}
\label{drops}

To explore the locality vs. non-locality issues of the plastic deformation we turn now to a direct exploration of the plastic
failure when our system is subjected to a simple shear. In such simulations one expects to see an elastic branch where the stress
increases linearly with the strain until the yield-stress is achieved and plastic events begin. There are a number of ways
that such simulations can be done. One way is by solving the so-called SLLOD equations \cite{SLLOD}, another way is to introduce and move two walls with
no-slip boundary conditions \cite{04VBB}, and the third is to impose small strain increments as described below and then to minimize the energy of the resulting configuration under the constraints imposed by the strain increment \cite{06ML}. We opt for the third option in order to be
able to accurately measure the distribution of stress and energy drops in the plastic events which are well defined only in this method. In doing so we will be able also to validate the results of \cite{06ML} and even to strengthen them.

\begin{figure}
\hspace {-1.1cm}
\centering
\includegraphics[scale = 0.8]{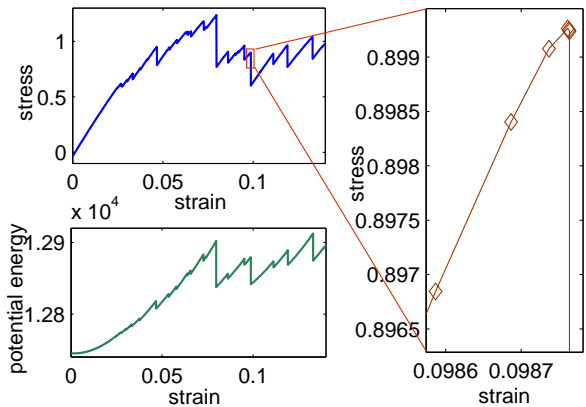}
\caption{ Typical stress-strain and potential energy-strain curves for a system with 4096 particles. The blow-up at the right panel demonstrates the procedure of reducing the increment steps to increase numerical precision in determining the stress and energy
drops.}
\label{SS}
\end{figure}

In detail, we prepare the same system described in Subsect. \ref{model} but now with $N=625$, 1024, $2500$, $4096$, $10000$ and $20164$, cooled at a rate of $10^{-3}$ $\epsilon/(k_B\tau_0)$. Beginning from a quenched unstrained configuration we impose a simple shear strain increment which will be denoted $\delta\epsilon$. This is achieved by applying the following transformation on the particles coordinates:
\begin{eqnarray}
r_x & \rightarrow & r_x + r_y\delta\epsilon\ , \nonumber\\
r_y & \rightarrow & r_y \ ,
\label{shear}
\end{eqnarray}
in addition to imposing Lees-Edwards boundary conditions. Typical stress-strain and potential energy-strain curves for a system
with $N=4096$ are shown in Fig. \ref{SS}. One sees the main elastic branch after which the evolution consists of small elastic branches ending with plastic drops. We employ a basic strain increment of $10^{-4}$ for systems
smaller than $N=10000$ and $5\times 10^{-5}$ for the larger systems. To increase our precision in determining the stress and energy drops and to guarantee that we do not overshoot and miss the next minimum we stop the simulation after a drop is detected, backtrack to the configuration prior to the drop and half the strain increments. The way the detection of the plastic drop is done is explained in detail in Appendix \ref{dropdet}. This procedure is repeated until the strain increment is smaller than $10^{-6}$ for systems smaller than $N=10000$ and $5\times 10^{-7}$ for the larger systems. An example of a trajectory approaching the plastic drop is shown in the blown-up right panel of Fig. \ref{SS}.
\subsection{System size dependence of stress and energy drops}
The statistics of the stress and energy drops is collected from between 4500 and 9000 plastic drops for each system size, where all the considered
drops are after steady state plastic flow had been reached, with measurements collected after about 40\% strains.
\begin{figure}
\hskip -1.6 cm
\centering
\includegraphics[scale = 0.6]{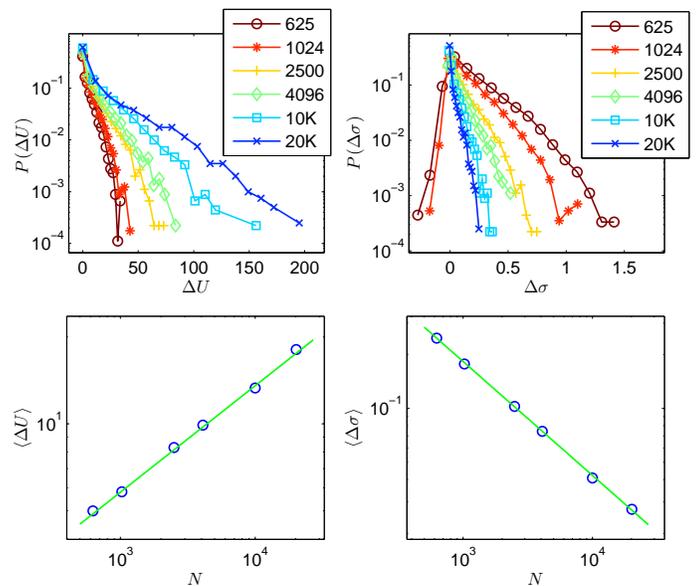}
\caption{Color online. The raw distributions of energy drops (upper left) and stress drops (upper right). The lower panels exhibit the system-size dependence of the mean energy drop (lower left) and mean stress drop (lower right). This system size dependence appears very accurately to conform with power laws $\langle \Delta U\rangle \sim N^{\alpha}$ and $\langle \Delta \sigma \rangle \sim N^{\beta}$ with $\alpha\approx 0.37$ and $\beta\approx-0.63$. These scaling laws are used to rescale the distributions as shown
in Fig. \ref{rescaled}}
\label{rawandmean}
\end{figure}

The raw distributions of the energy and stress drops are displayed in the upper panels of Fig. \ref{rawandmean}. The lower panel of the same figure display the system size dependence of the average energy drop and average stress drop respectively, in a log-log plot. The conclusion is that the mean drop are described to a very high precision by power laws of the form
\begin{equation}
\langle \Delta U\rangle \sim N^{\alpha}, \quad \langle \Delta \sigma \rangle \sim N^{\beta}  \ ,
\label{laws}
\end{equation}
with
\begin{equation}
\boxed{\alpha\approx 0.37, \quad\beta\approx-0.63} \ .
\end{equation}
While we did not expect these scaling exponents, it is very easy to understand that there exists a scaling relation between them,
\begin{equation}
\alpha-\beta=1 \ .
\label{screl}
\end{equation}
To see this, we note that in the athermal limit at very low strain rates the steady state plastic flow occurs around a fixed value of the stress, which is the yield stress $\sigma_Y$. On the average, for every elastic increase in the energy which occurs
for a strain increment $\Delta \epsilon$ we have a corresponding plastic drop $\Delta \sigma$, and we can estimate the mean energy drop according to
\begin{equation}
\sigma_Y \times\langle \Delta\epsilon \rangle \times V = \sigma_Y\times \frac{\langle\Delta \sigma\rangle }{\mu}\times  V =\langle \Delta U \rangle
\end{equation}
Since in our systems of fixed density $V\sim N$, the scaling relation (\ref{screl}) follows immediately.

\begin{figure}
\hskip -1.2cm
\centering
\includegraphics[scale = 0.48]{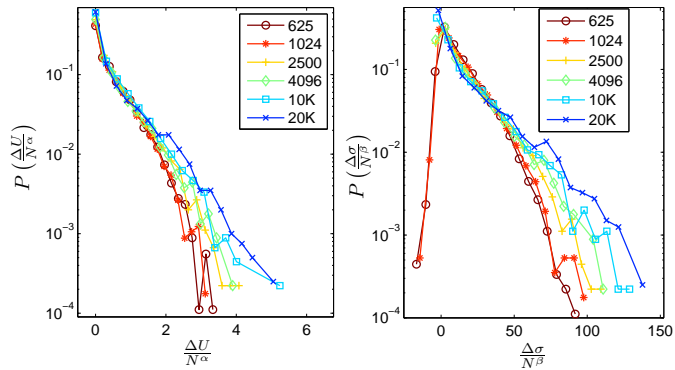}
\caption{Color online. The distributions of energy drops (left panel) and stress drops (right panel) shown in the upper panels of Fig. \ref{SS} after rescaling by energy drops by $N^\alpha$ and the stress drops by $N^\beta$. Note that the data collapse is quite perfect for probabilities larger than about 1\%, and then begin to meander systematically; we cannot determine at this point whether this meandering is due to paucity of data or due to multiscaling of the higher moments. }
\label{rescaled}
\end{figure}

We can use the scaling laws Eq. (\ref{laws}) to re-scale the raw distribution functions of Fig. \ref{rawandmean}. The resulting distributions are presented in Fig. \ref{rescaled}. We note that the data collapse is superb for probabilities larger
than about 1\%. For lower probabilities the quality of the collapse deteriorates;  due to the paucity of data there we cannot
determine whether the deteriorating collapse is due to multiscaling of higher moments or due to statistical errors. If we attempt to rescale using a somewhat higher exponents we can get the tails to collapse but then the high probabilities fail to collapse.
The scaling Eq. (\ref{laws}) and the quality of the data collapse demonstrate that the plastic drops are not localized, but in fact are sub-extensive. This finding is in agreement with \cite{06ML,06TLB}, although the numerical value of our scaling exponents $\alpha$
and $\beta$ differs from theirs (both \cite{06ML} and \cite{06TLB} report $\alpha=-\beta=0.5$). We return to the issue of
non-universality of the scaling exponents in Sect. \ref{another}.

\begin{figure}
\hskip -1.0cm
\centering
\includegraphics[scale = 0.55]{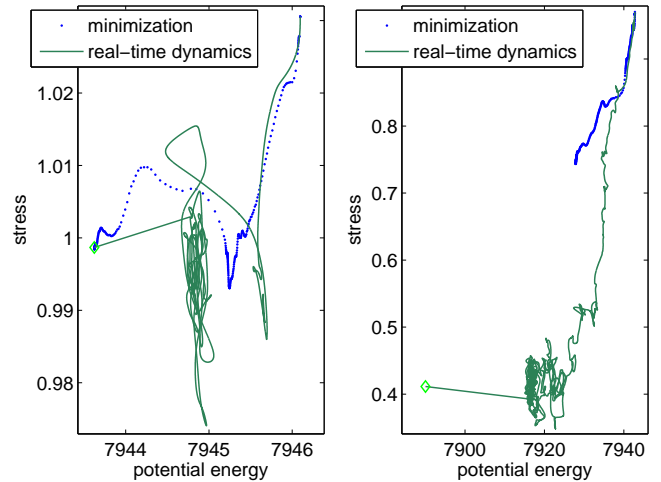}
\caption{Color online. Left panel: an example for which the initial and final energy and stress values are the same
for the energy minimization algorithm and the real-time dynamics. In dotted blue line is the latter trajectory and
in continuous green line is the former. For the dynamics the green diamond represents the quenched configuration after
the trajectory had reached steady state. Right panel: and example for which the true real-time dynamics results in a bigger
and energy and stress drops compared to the energy minimization procedure. Both panels are for systems with $N=2500$ in the
steady state plastic flow. }
\label{dynamics}
\end{figure}

As mentioned in the introduction, the work of \cite{06ML} was criticized on the basis of their algorithm; it was proposed
that the energy minimization procedure follows an un-physical trajectory in configuration space, and increases artificially
the amount of energy drop in a plastic event. Also, since the energy minimization does not correspond to real time units,
the system has effectively infinite time to reach the minimum energy, and this this is not a proper small strain rate limit.
We demonstrate now that this criticism is incorrect, and in fact molecular dynamics can often lead to {\bf larger} energy drops.
Energy minimization often stops in the next available minimum, whereas true dynamics can often trigger subsequent energy drops
and ends up increasing the size of the average region involved in the plastic event. This conclusion is demonstrated with
trajectories of the two methods in Fig. \ref{dynamics}, and a movie of a multiple avalanche that is seen in true molecular
dynamics is available in \cite{web}.
We conclude that the energy minimization procedure produces a {\em lower bound} to the energy drops rather than an exaggerated
result. One should understand that the true dynamics releases kinetic energy after the first
drop which is negligible on the system scale, being the difference between a single saddle and a neighboring minimum. But before
this negligible energy is spread over the whole system it heats up considerably the local neighborhood and can easily trigger
further energy drops which in turn heat up the local region even further. Accordingly the `energy drop' is well defined
only within the energy minimization procedure, where it is not in the eyes of the beholder.

One could argue that the reason that the energy drops are sub-extensive is only because every change in an elasto-plastic medium involves also an elastic relaxation. This is of course true;
since elasticity is long-ranged, one could expect some kind of sub-extensive scaling. In fact, this underlines
our belief that it is futile to talk about pure plastic energy changes, since the contribution of the purely plastic
part of an energy drop can be very small compared to the associated elastic contribution. Furthermore, any measure that employs the coordinates of all the particles in the system, like various participation numbers, will always collect some elastic
relaxation contributions which can be quite large. Notwithstanding, we propose here that the sub-extensive scaling of the
energy drops is not {\bf only} because of the elastic contributions, and that we can demonstrate this scaling also when
we carefully exclude the elastic contributions. This will be done in the next subsection.

\subsection{System-size dependence of purely plastic contributions}
\label{pure}
\begin{figure}
\hskip -1.2cm
\centering
\includegraphics[scale = 0.48]{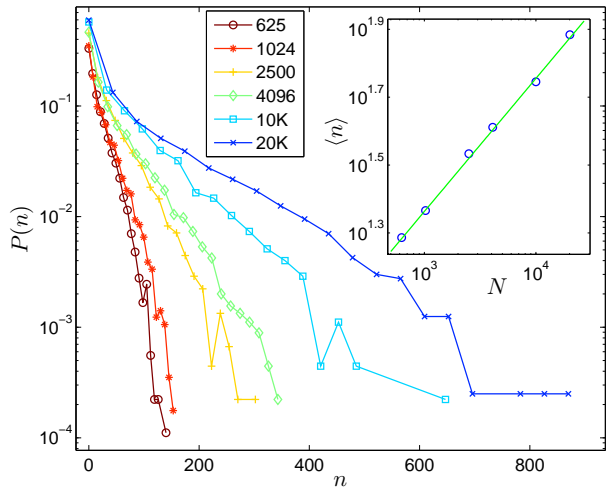}
\caption{The distributions of $n$ for different size systems. Inset: the scaling of the average of $n$ with the
system size. A power law is detected, see text. }
\label{distance}
\end{figure}

To single out the purely plastic contributions to the elasto-plastic events we will track the neighbor lists during the straining simulations. Before every flow event each particle in the system is assigned a neighbor list consisting of the particles residing within the range of interaction, cf. Eq. (\ref{potential}). After the elasto-plastic event each particle is checked against the original neighbor list and the number of neighbor changes is monitored. It is important to state that upon loading, the neighbor list may change due to non-affine elastic effects when a neighboring particle
leaves the range of interaction or a new particle enters that range. During an affine linear elastic loading the neighbor list does not change at all. However, during the loading we encounter also nonlinear non-affine elasticity. We have checked carefully and determined that during the latter there exist T1 processes in which the neighbor list does change. Such processes are reversible
and can be traced back by unloading. In our simulations we find that in order to detect more than one change in the neighbor list per particle in the non-affine elastic processes we need to undergo very large strain intervals. We therefore conclude that during the plastic drops, where the displacement field changes only little, it is unlikely that the non-affine elastic strains would involve more than one change per particle in the neighbor list. We thus propose that by
filtering those particles whose neighbor list had undergone more than one change during the elasto-plastic event we capture essentially only those that were a part of a purely plastic irreversible event. We formed a measure of the size $n$ of the purely plastic event from summing up the
number of particles whose neighbor list changed by more than unity. In Fig. \ref{distance} we show the
raw distribution of $n$ for systems of varying sizes, with an inset exhibiting the average of $n$.
As before, we find that the raw distribution tend to higher and higher values of $n$ when the system size increases, and
as before we find that the average of $n$ scales nicely with the system size,
\begin{equation}
\langle n \rangle \sim N^\zeta \ ,
\end{equation}
with $\zeta\approx 0.39$. Note that with the present accuracy we cannot rule out that $\zeta=\alpha$.

\section{Yet another model to remove the last doubts}
\label{another}

\begin{figure}
\hskip -1.2cm
\centering
\includegraphics[scale = 0.55]{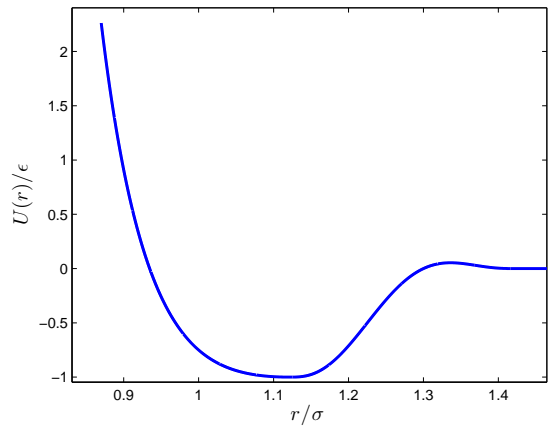}
\caption{The potential chosen for the model of section \ref{another}. }
\label{poten2}
\end{figure}

The purpose of this section is to remove the last doubts about the sub-extensivity of the plastic events.
To this aim we introduce a new model which has a measure of plastic deformation
that is insensitive {\em by construction} to non-affine elasticity. As above the system consists of
point particles in two-dimensions interacting via a pair-wise potential
\begin{equation}\label{potent2}
U(r) =
\left\{
\begin{array}{l}
\epsilon\left[\left(\frac{\sigma}{r}\right)^{12} -
\left(\frac{\sigma}{r}\right)^6 + \frac{1}{4} - h_0 \right]  \ ,\quad  r \le \sigma x_0 \\
\epsilon h_0P\left(\frac{\frac{r}{\sigma} - x_0}{x_c}\right)  \ , \quad \sigma x_0 < r  \le \sigma (x_0 + x_c)  \\
0 \ , \quad  r > \sigma(x_0 + x_c) \ ,
\end{array}
\right.
\end{equation}
which consists of a shifted repulsive part
of the standard Lennard-Jones potential, connected via a hump to a region that is smoothed continuously to zero
(up to second derivatives), cf Fig. \ref{poten2}.
The point $x_0$ is the position at which the LJ potential is minimal, $x_0 \equiv 2^{1/6}$, and the position where the
potential vanishes is $\sigma(x_0+x_c)$. The parameter $h_0$ determines the depth of the minimum. The polynomial $P(x)$ is chosen
as \begin{equation}\label{defineP}
P(x) = \sum_{i=0}^{6}A_i x^i\ .
\end{equation}
with the coefficients given in table \ref{table}. Note that with these parameters the position of the shallow maximum
in Fig. \ref{poten2} is $r_b = 1.336~189~578~406~025$.

\begin{table}
\begin{tabular}{|c|c|}
\hline
$A_0$&-1.0\\
$A_1$&0.\\
$A_2$&0.642897426047121\\
$A_3$&30.503000042685606\\
$A_4$&-80.366384684339579\\
$A_5$&72.652179536433835\\
$A_6$&-22.431692320826986\\
\hline
\end{tabular}
\caption{The coefficients in Eq. (\ref{defineP})}
\label{table}
\end{table}

We use the position of $r_b$ to define events that are plastic by definition and not elastic: whenever the distance between a pair of particles which were bonded exceeds $r_b$ their energy drops irreversibly and spreads around. Similarly, whenever a pair
of particles that were not bonded forms a new bond when their distance becomes smaller than $r_b$ their energy drops irreversibly and is spread around. We can thus simply count the number of particles that underwent a bond break or a bond creation during a stress drop, to obtain an unquestionable measure of the size of the purely plastic event. To do this, we repeat the straining experiment in much the same way as discussed above, but for systems with the present potential (\ref{potent2}) for system sizes of
$N=1024$, 2500, 4096 and 10 000, keeping the same value of the density $\rho$ as before. In every stress drop event we count the number of particles experiencing a bond change, which is denoted as $n_b$. We stress that we carefully ascertained that there is
absolutely no change in this measure except during the drops, even when the system undergoes very substantial non-affine elastic displacements.
\begin{figure}
\hskip -1.2cm
\centering
\includegraphics[scale = 0.55]{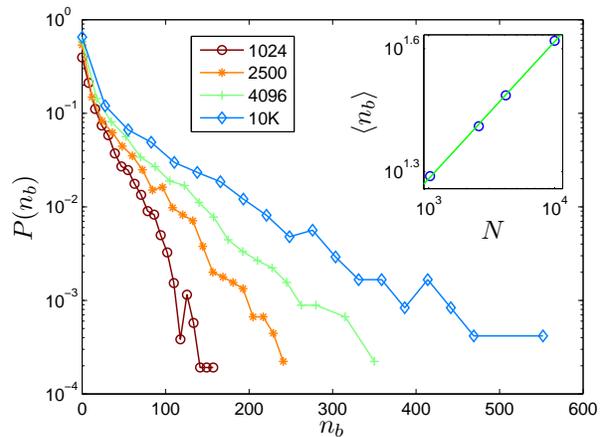}
\caption{The distribution of the number of particles participating in a purely plastic event as a function of the system
size. The number $n_b$ is defined such that elastic processes cannot contribute to this measure by construction. In the inset we show a log-log plot of the average $\langle n_b\rangle$ as a function of $N$; the scaling law is Eq. (\ref{defchi}). }
\label{dist}
\end{figure}
Fig. \ref{dist} displays the distribution of $n_b$ as a function of system size and also in the inset the dependence of $\langle n_b \rangle$ as a function of $N$ in a log-log plot. The best fit reads
\begin{equation}
\langle n_b \rangle \sim N^\chi \ ,
\label{defchi}
\end{equation}
with $\chi\approx 0.33$. We believe that the difference between $\chi$ and $\zeta$ is outside the error bars for either number, indicating that changing the potential very well may result in changing the value of the scaling exponents.

\section{Summary and conclusions}
\label{summary}

We have examined the issue of the non-locality of plastic deformation by a set of numerical experiments of increasing stringency in filtering out the elastic contributions. First we examined the maximal force exerted on a particle moving within a glassy medium of finite size, and discovered that the average maximal force before a plastic deformation depends on the system size as a power law. Not being able to filter out the elastic from the plastic effects in this experiment, we turned to examining the distribution of stress drops and energy drops in a straining experiment once the system exceeded the yield stress and landed on the steady state plastic flow. Again we found that both the stress and the energy drops had distribution function that exhibited an interesting scaling law with the size of the system, strengthening the conclusion that the plastic events are not localized. The scaling exponents found were non trivial, but in agreement with the presumably exact scaling relation Eq. (\ref{screl}). Having still difficulties in distinguishing elastic from plastic contributions in these measurements, we turned to the neighbor lists whose large changes are most likely not due to non-affine elasticity. Those changes scaled again with the size of the system, and the scaling exponent $\zeta$ was numerically sufficiently close to $\alpha$ to indicate that they are the same exponents, and that the scaling of the energy drops in the plastic events is the same as that of the purely plastic contribution. Finally, to remove the last doubts we constructed a model in which the plastic events can be accurately separated from any elastic contribution, and found that also there the size of the plastic events scales sub-extensively with the size of the system. The exponent $\chi$ was sufficiently different from $\alpha$ or $\zeta$ to indicate a lack of universality in these exponents. It is very likely that the scaling exponents seen here depend on the details of the models, on the nature of disorder and on parameters like pressure, density etc. We propose that understanding these exponents and finding a theoretical calculation of them will shed important light on the
physics of amorphous solids.

Finally, we remind the reader that the experiments reported here are done at zero temperature and at infinite heat extraction rates. The situation at finite temperature and finite strain rates needs a separate examination which is beyond the scope of this paper. In \cite{05TD} the magnitude of stress drops in strained bubble rafts were found to depend very weakly, if at all, on number of bubbles in the raft. Whether this is due to the finite strain rate or due to some salient different physics remains to be answered by future careful simulations of the type presented above but with finite strain rates.

\acknowledgments

Part of this work was executed while both authors visited the Institute of Theoretical Physics at the Chinese University of Hong Kong. We thank the Institute and in particular Emily S.C. Ching for their hospitality. This work has been supported in part by the German Israeli Foundation, the Israel Science Foundation and the Minerva Foundation, Munich, Germany. Useful discussions with Eran Bouchbinder and George Hentschel are acknowledged.

\appendix

\section{The detection of plastic drops}
\label{dropdet}

\begin{figure}[ht]
\centering
\includegraphics[scale = 0.55]{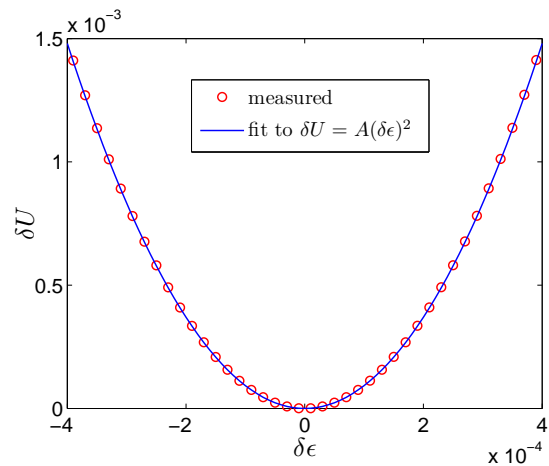}
\caption{The measured (red dots) energy difference $\delta U \equiv U_{\rm aff} - U_0$ as a function of
the strain increment $\delta\epsilon$, and the fit (blue continuous line) to $\delta U = A(\delta\epsilon)^2$,
with $A = 9285.0$.}
\label{symmetric}
\end{figure}

Since our system is disordered, the affine transformation defined in (\ref{shear})
doesn't take the system to a minimum of the potential under the strain constraint.
Instead, the system reaches some state having the energy $U_{\rm aff}$.
We denote the energy read after minimizing this state as $U_0$,
and define the difference between the energies of the system after imposing
the affine strain increment and the resulting minimized configuration as
$\delta U \equiv U_{\rm aff} - U_0$.
We expect this difference to be proportional to the system size, since
it must be extensive. Also, since this quantity represents the extent to which the affine
transformation doesn't match the new potential energy minimum,
we expect this mismatch to be symmetric under the direction in which the strain
increment is imposed starting from a given configuration. So, these considerations lead us to
assume $\delta U \sim N(\delta\epsilon)^2$.
In Fig. \ref{symmetric} the measured function $\delta U(\delta\epsilon)$ is given for some
initial state randomly selected from the obtained steady state
configurations of the $N = 1024$ particle system.

\begin{figure}[ht]
\centering
\includegraphics[scale = 0.55]{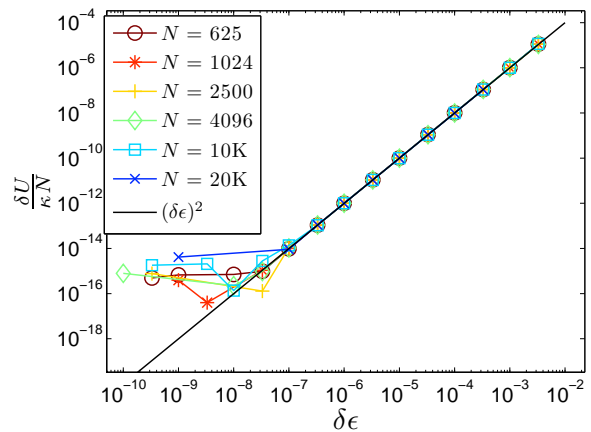}
\caption{The scaled energy difference $\frac{\delta U}{\kappa N}$ (symbols), compared to
$\frac{\delta U}{\kappa N} = (\delta\epsilon)^2$ (black continuous line).
Double precision numerics allows us to rely on this quantity only up to strain increments
of $\delta\epsilon \sim 10^{-7}$, below which large errors are accumulated.}
\label{loglogDeltaU}
\end{figure}

The functional form of $\delta U(\delta\epsilon)$ is indeed symmetric and follows
a quadratic form with good precision up to strain increments well above the
basic strain increment step used in our simulations. To check how far down
in strain increment step we can go using this measure,
we repeat a similar measurement, now over some orders of magnitude in
$\delta\epsilon$, and for all simulated system sizes. The results are displayed in
Fig. \ref{loglogDeltaU}; it is obvious that double-precision
numerics allow us to rely on $\delta U$ measured by potential energy differences only up to strain increments
of $\delta\epsilon > 10^{-7}$, below which large errors are accumulated.
This limitation has already been discussed in length in \cite{barrat} for a similar experiment.
The energy differences in Fig. \ref{loglogDeltaU} are rescaled by the system size $N$ and
the factor $\kappa \equiv \frac{\delta U}{N(\delta\epsilon)^2}$
measured in the $\delta\epsilon$ range in which measurements
of $\delta U$ are reliable. For our system $\kappa \sim 10.0$ far from instabilities.

\begin{figure}
\centering
\includegraphics[scale = 0.6]{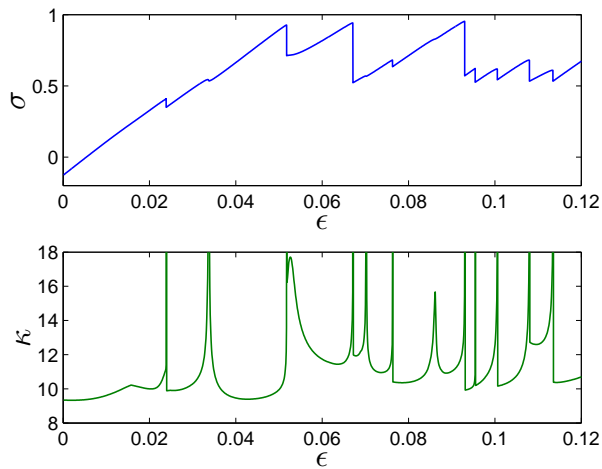}
\caption{Demonstration of the qualitative behavior of $\kappa$ (lower panel, see text),
in relation with the plastic stress drops (upper panel), for a typical
$N = 1024$ run approaching steady state flow.}
\label{kappaEvolution}
\end{figure}

In Fig. \ref{kappaEvolution} we display the evolution of $\kappa$ throughout
a typical simulation run for a system of $N=1024$ particles.
The divergences of $\kappa$ are completely analogous to
those analyzed in great detail in \cite{06ML} for a different (but closely related)
set of quantities, where a $\sqrt{\epsilon_c - \epsilon}$ law was
found. It turns out that for the strain increments used in
our experiments $\kappa$ is bounded from above by $\sim 10^2$,
unless a plastic flow event has occurred. In the latter case,
the measured value of $\kappa$ averages around $\sim 10^9$, for all system sizes.
This enormous difference in orders of magnitude of
$\kappa$ makes it an extremely robust measure
for detection of plastic events.

\begin{figure}
\centering
\includegraphics[scale = 0.55]{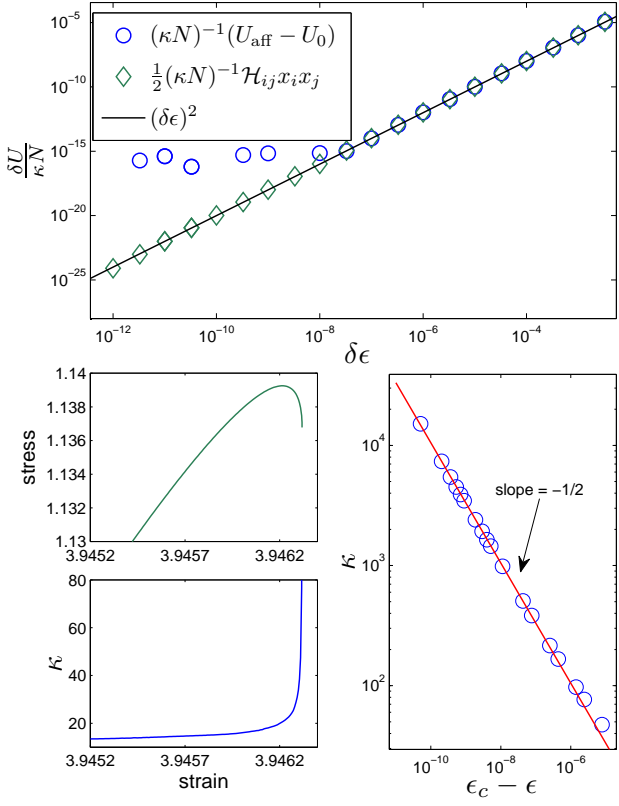}
\caption{Top panel: comparison of $\delta U$ calculated by energy differences
(blue circles) to the same quantity calculated via (\ref{energyDifference})
(green diamonds). The black line
corresponds to $\frac{\delta U}{\kappa N} = (\delta\epsilon)^2$.
Using the hessian allows us to go to $\delta\epsilon$ values that are orders of magnitude
smaller than those enabled by the energy difference.
Lower panels: analysis of the onset of a typical flow event for a $N=625$ system;
the lower left panels show the stress-strain and the $\kappa$-strain curves.
The lower right panel demonstrates the $\frac{1}{\sqrt{\epsilon_0 - \epsilon}}$
divergence, in qualitative agreement with \cite{06ML}. The red continuous line has
a slope of --1/2.}
\label{divergence}
\end{figure}

To better establish the connection with the findings in \cite{06ML}, we
first note that while $\delta U$ cannot be calculated to the required precision
by computing differences, we can improve the accuracy in the $\delta U$
measurements using an equivalent calculation.
We impose the transformation (\ref{shear}), and calculate the
displacement vector $\delta x_i$ for each particle, after minimizing the transformed state.
Using the hessian matrix definition
${\cal H}_{ij} \equiv \frac{\partial^2 U}{\partial x_i \partial x_j}$, we can expand the
potential energy near the minimum,
\begin{equation}\label{energyDifference}
U(\{\delta x\}) - U_0  \simeq  \frac{1}{2}{\cal H}_{ij}\delta x_i \delta x_j\ ,
\end{equation}
The LHS of (\ref{energyDifference}) is the exact definition of $\delta U$,
but as opposed to the energy difference calculation, it consists of a sum of terms
of the same order, which takes advantage of the availability of an analytic expression
for ${\cal H}_{ij}$. In the top panel of Fig. \ref{divergence} the improvement of
the use of (\ref{energyDifference}) is apparent, which verifies that this measure can
be used to validate the agreement with \cite{06ML}. A description of the onset of a
typical plastic event for a $N=625$ system is displayed in the bottom panels
of Fig. \ref{divergence}, where a similar $(\sqrt{\epsilon_0 - \epsilon})^{-1}$ divergence
is found for the quantity $\kappa$.

\end{document}